\documentclass[10pt,letterpaper,twocolumn]{article} %% two column, final layout

\usepackage{ol}
\usepackage{hyperref}
\usepackage{amsmath}

\begin{document}

\twocolumn[ %% activate for two-column option

\title{Design of switches and beam splitters using chaotic cavities}

\author{O. Bendix and J. A. M\'endez-Berm\'udez}

\address{Max-Planck-Institut f\"ur Dynamik und Selbstorganisation, Bunsenstra\ss e 10,
D-37073 G\"ottingen, Germany}

% Do not use \email or \homepage here. E-mail and URL can be given just before references.

\begin{abstract}
We propose the construction of electromagnetic (or electronic)
switches and beam
splitters using chaotic two-dimensional multi-port waveguides.
A prototype two-port waveguide is locally deformed in order to
produce a ternary incomplete horseshoe proper of mixed phase space
(chaotic regions surrounding islands of stability where motion is regular).
Due to tunneling to the phase space stability islands
the appearance of quasi-bound states (QBS) is induced.
Then, we attach transversal ports to the waveguide on the
deformation region in positions where the phase space structure
is only slightly perturbed.
We show how QBS can be guided out of the waveguide through
the attached transversal ports giving rise to frequency
selective switches and beam splitters.
\end{abstract}


] %% activate for two-column option

\noindent Switches and beam splitters are key elements in
optical information processing, imaging, integrated photonic, and
optical communication systems.
In this letter we propose a novel way to construct
switches and beam splitters using two-dimensional (2D) chaotic
waveguides.

\begin{figure}[htb]
\centerline{\includegraphics[width=8cm]{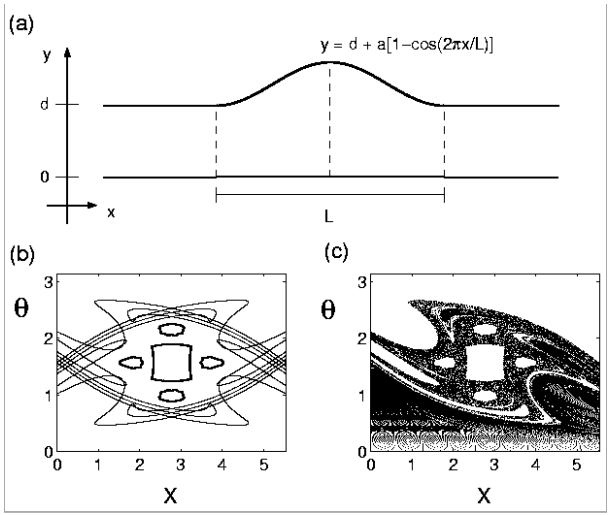}}
\caption{(a) Geometry of the waveguide, (b) horseshoe, and (c)
transient Poincar\`{e} map with surface of section at $y=0$, $(d,a,L)=(1.0,0.305,5.55)$.
The tendrils\cite{horseshoes} of the horseshoe in (b) are plotted up to level three. The
thick lines in (b) are the limits of stability islands. The transient
Poincar\`{e} maps in (c) is constructed for rays entering the cavity from the left
port.}
\label{fig:wguide}
\end{figure}

The 2D waveguide we shall use is formed by a cavity connected
to two collinear semi-infinite ports of width $d$ extended along the $x$-axis. The
prototype cavity has the geometry of the so-called {\it cosine
billiard}\cite{infinite,infinite1,finite2,finite1}: it has a flat wall at
$y=0$ and a deformed wall given by $y(x)=d+a [1-\cos(2\pi x/L)]$, where $a$ is the
amplitude of the deformation and $L$ is the length of the cavity. In Fig.
\ref{fig:wguide}(a) we show the geometry of the waveguide.

In order to obtain the panorama of the ray (classical) dynamics of
the waveguide we construct its Smale horseshoe\cite{horseshoes}.
It gives the topology of the homoclinic tangle which completely
characterizes the scattering dynamics, connecting the interacting region with
the asymptotic regions.
For our waveguide, the domain of the interacting region
is the cavity, while the asymptotic regions are the ports. The number of fundamental
orbits (period-one periodic orbits) determine the order of the horseshoe. Our cavity
(for $a>0$) has three of them (shown in dashed lines in Fig. \ref{fig:wguide}(a)) and
hence its horseshoe is ternary. The horseshoe is formed by
the invariant manifolds (stable and unstable) of the hyperbolic fixed points of the
cavity (the ones located at the edges of the cavity).
We plot the horseshoe using a Poincar\`{e} Map\cite{Licht} (PM), i.e., we follow
orbits with initial conditions along the manifolds and each time a ray
impinges on the flat wall of the waveguide (chosen as surface of section)
we plot the position $x$ and the angle $\theta$, the angle the ray makes with the
lower boundary.
In Fig. \ref{fig:wguide}(b) we present the horseshoe of the waveguide with parameters
$(d,a,L) = (1.0,0.305,5.55)$ where only the tendrils\cite{horseshoes} up to the
hierarchy level three
are plotted. In particular for this set of parameters the horseshoe is incomplete;
a typical situation of mixed phase space.

For the set of parameters used here, the cavity develops a period-one and
period-four resonance islands whose boundaries are shown with thick lines
in Fig. \ref{fig:wguide}(b). These islands are formed by trapped
orbits bouncing in the neighborhood of stable periodic orbits. In particular the
central island is formed by trajectories colliding nearly perpendicular with
the walls around $x=L/2$.
Note that the orbits within these islands are not accessible to scattering
rays.

To further analyze our waveguide system we use the {\it transient Poincar\`{e}
map}\cite{finite2} which is a PM where the initial conditions are chosen to lie
outside the cavity. The transient PM
generated by trajectories whose initial conditions start
in the left port is presented in Fig. \ref{fig:wguide}(c).
Comparison of Fig. \ref{fig:wguide}(b) and Fig.
\ref{fig:wguide}(c) shows, as expected, that the stability islands produced by bounded
motion inside the cavity are forbidden phase space regions for scattering trajectories.
Also note that the structure of the transient PM shadows the unstable manifold of the
corresponding horseshoe.

To study the wave scattering phenomena we numerically solve the
Schr\"{o}dinger equation to compute the scattering wave functions
(by means of the Fisher-Lee relation\cite{datta}) and the
energy-dependent conductance $G_{pq}$ (using the Landauer-B\"{u}ttiker
formalism\cite{datta}), measured from port $q$ to port $p$.
We want to emphasize that since the problem of a quantum wave in a 2D billiard
(Schr\"{o}dinger equation) is equivalent to that of a TM wave inside a 2D waveguide
with Dirichlet boundary conditions (Helmholtz equation)\cite{stock},
our analysis is applicable to both, electronic and electromagnetic setups.

Additionally, to explore ray-wave correspondence we use the Husimi
distribution\cite{hus} which is the projection of a given state onto
a coherent state of minimum uncertainty. The Husimi distribution is
a phase space probability density that can be directly compared to the
ray phase space.

\begin{figure}[t]
\centerline{\includegraphics[width=8cm]{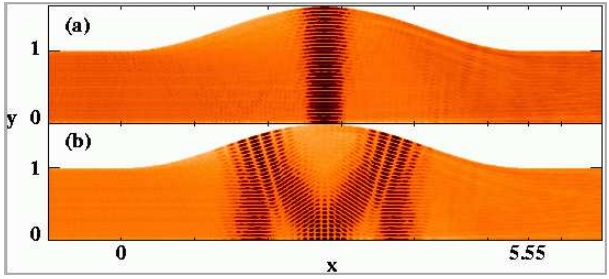}}
\caption{Density plots for scattering wave functions calculated at resonance
energies (light-dark = minimum-maximum).
(a) $E=4008.8583$, (b) $E=4027.0597$.
These QBS are constructed by averaging all the ${\cal M}=20$ open modes.
$E=2d^2E_{\mbox{\tiny W}} \sim {\cal M}^2\pi^2$; where $E_{\mbox{\tiny W}}$
is the energy of the wave and ${\cal M}$ is the largest
transversal mode $m$ beyond which the longitudinal wave vector
$[2E_{\mbox{\tiny W}}-(m\pi/d)^2]^{1/2}$ becomes complex.}
\label{fig:QBS}
\end{figure}

It has been shown\cite{finite2,finite1} that when the cavity of the waveguide
of Fig. \ref{fig:wguide}(a) is characterized by an incomplete horseshoe the
conductance $G_{\mbox{\tiny LR}}$ (= $G_{\mbox{\tiny RL}}$)
fluctuates strongly with sharp resonances.
The wave functions belonging to the {\it sharpest}
conductance resonances can be identified with energy eigenstates living in the phase
space stability islands. In particular, we noticed\cite{lass2}
that some wave functions at resonance
reveal I-, V- and M-shaped patterns which shadow the ray trajectory of a
particle in a period one, two and four periodic orbits, respectively.
The reason of this phenomenon is Heisenberg's uncertainty principle that allows
scattering wave functions to tunnel through KAM barriers\cite{Licht}.
Since the scattering wave functions that tunnel into classical islands are very
similar to a set of eigenfunctions of the corresponding closed cavity, we
named them {\it quasi bound states} (QBS). For the set of parameters chosen for
the cavity in this work, the QBS may be of two types only: I-type or M-type.
See the corresponding scattering wave functions in Fig. \ref{fig:QBS}
found at $G_{\mbox{\tiny LR}}$ local minima for waves incident from the
left port.
Notice that under resonance conditions the cavity acts as a resonator.
The I-type (M-type) QBS have support on
the period-one (period-four) resonance island, as can be clearly seen
in Fig. \ref{fig:hus}(a-b) where we plot the Husimi distributions for
the QBS of Fig. \ref{fig:QBS} together with the transient PM of the system.

\begin{figure}[t]
\centerline{\includegraphics[width=6.5cm]{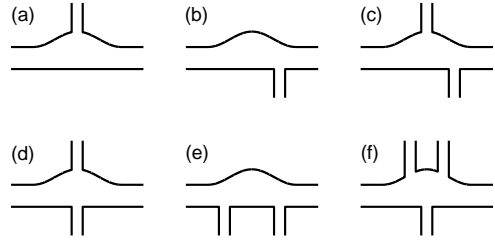}}
\caption{Examples of (a-c) switch and (d-f) beam splitter setups.}
\label{fig:setups}
\end{figure}

\begin{figure}[t]
\centerline{\includegraphics[width=8cm]{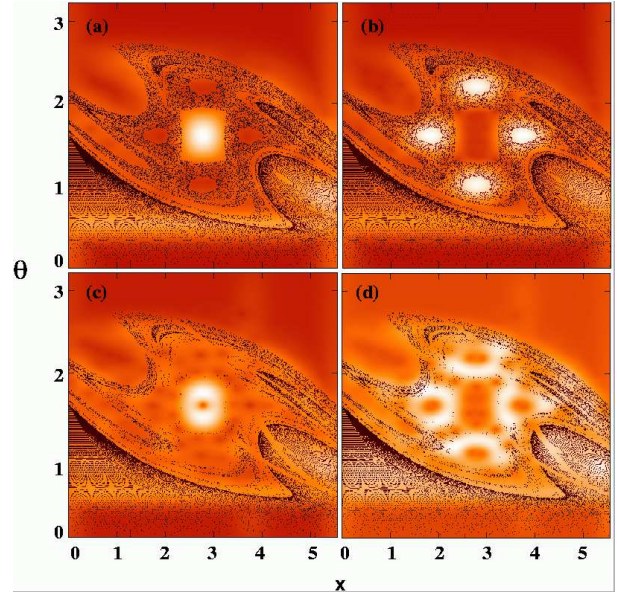}}
\caption{Transient Poincar\`{e} maps (dots) and Husimi distributions
(orange-white = minimum-maximum) with surface of section at $y=0$
constructed for rays/waves entering the waveguide of (a-b) Fig.
\ref{fig:wguide}(a) and (c-d) Fig. \ref{fig:setups}(c) from the left port.
In (a) and (b) the Husimi distributions for the QBS of Fig. \ref{fig:QBS}(a)
and Fig. \ref{fig:QBS}(b) are shown, respectively.
In (c) and (d) the Husimi distributions for
scattering sates at $G_{\mbox{\tiny LU}}$ and $G_{\mbox{\tiny LD}}$ maxima
are plotted, respectively, both at $G_{\mbox{\tiny LR}}$ minima.
(c) $E=3997.34$, (d) $E=4103.4096$.}
\label{fig:hus}
\end{figure}

Now, the idea is to attach transversal ports to the cavity
in order to guide the QBS out of the waveguide
(once they are excited by the appropriate resonance energy)
to construct switches and/or beam splitters.
Without loss of generality, we will attach semi-infinite ports
of width $d_{\mbox{\tiny T}}$ extended along the $y$-axis
(see some multi-port setups in Fig. \ref{fig:setups}).
Then, to make use of the I-type
QBS the ports have to be attached to the waveguide centered at $x=L/2$
(the position of the period-one periodic orbit supporting I-type QBS)
on the lower boundary, on the upper one [Fig. \ref{fig:setups}(a)],
or on both [Fig. \ref{fig:setups}(d)].
While to use the M-type QBS, the ports should be centered at $x=L/2$ and
$L/2 \pm 0.975$ on the lower boundary or/and at $x = L/2 \pm 0.975$
on the upper one, see for example Figs. \ref{fig:setups}(b,e,f).
In particular we will use the setup of Fig. \ref{fig:setups}(c) which
combines both, I- and M-type QBS.

\begin{figure}[ht]
\centerline{\includegraphics[width=8cm]{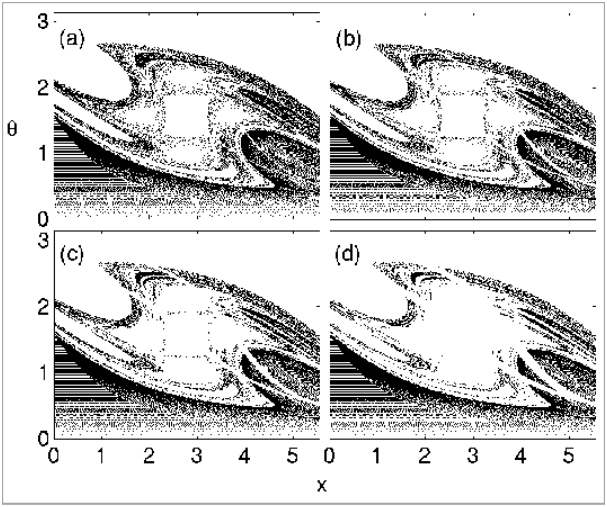}}
\caption{Transient Poincar\`{e} maps with surface of section at $y=0$
constructed for rays entering the beam splitter setup of Fig.
\ref{fig:setups}(c) from the left port. (a) $d_{\mbox{\tiny T}}=0.05$,
(b) $d_{\mbox{\tiny T}}=0.10$, (c) $d_{\mbox{\tiny T}}=0.15$,
and (d) $d_{\mbox{\tiny T}}=0.25$.}
\label{fig:tpm}
\end{figure}

Evidently, by attaching transversal ports to
the waveguide the cavity phase space is perturbed.
However since we want to make use of QBS, the width $d_{\mbox{\tiny T}}$
of the attached ports must be small enough to preserve the global phase
space structure.
Then, in Fig. \ref{fig:tpm} we analyze the effect of the size of the
transversal ports on the phase space for the setup of Fig. \ref{fig:setups}(c)
by plotting the transient PM for different values of $d_{\mbox{\tiny T}}$.
Notice that by increasing $d_{\mbox{\tiny T}}$, the chaotic region separating
the period-one and period-four stability islands becomes narrower until it
completely disappears.
Thus, we will use a value of $d_{\mbox{\tiny T}}$ for which the transient PM
still shows period-one and period-four stability regions:
$d_{\mbox{\tiny T}}=0.15$, see Fig. \ref{fig:tpm}(c).

Remember that the QBS shown in Fig. \ref{fig:QBS} are produced at
frequencies that correspond to sharp local minima in $G_{\mbox{\tiny LR}}$.
Now, for the multi-port setup of Fig. \ref{fig:setups}(c) we can
also measure the conductance from the left to the upper port,
$G_{\mbox{\tiny LU}}$, and from the left to the lower port,
$G_{\mbox{\tiny LD}}$.
We observe, as expected, that some of the sharpest minima in
$G_{\mbox{\tiny LR}}$
correspond to maxima in $G_{\mbox{\tiny LU}}$ or $G_{\mbox{\tiny LD}}$,
meaning that at such frequencies a wave tunneling to the stability
regions is guided up or down through the transversal ports.
To show such behavior,
in Fig. \ref{fig:hus} we compare the Husimi distributions for the QBS of
Fig. \ref{fig:QBS} (waveguide without transversal ports) to Husimi distributions
for scattering states at $G_{\mbox{\tiny LR}}$ resonance for the muti-port setup
Fig. \ref{fig:setups}(c).
The states of Figs. \ref{fig:hus}(c) and \ref{fig:hus}(d) correspond to minima in
$G_{\mbox{\tiny LR}}$ and maxima in $G_{\mbox{\tiny LU}}$ and $G_{\mbox{\tiny LD}}$,
respectively.
As expected since $G_{\mbox{\tiny LR}}$ is minimum, the maximum probability of
the Husimi distributions for these states is located inside phase space stability
regions (as the QBS of the two-port terminal do).
However, there is a probability minimum (equal to zero) at the center of the
stable regions indicating that in fact the QBS are guided out from the cavity
through the transversal ports.

Note that with the setup of Fig. \ref{fig:setups}(c) one is able to
guide a beam up (down) by exciting an I- (M-) type QBS, thus,
giving rise to a frequency selective switch.
While, a beam splitter
may be constructed by using the setup of Fig. \ref{fig:setups}(d)
(Fig. \ref{fig:setups}(e)), since an I- (M-) type QBS could be
{\it splitted} up and down (down) at certain resonant frequencies.
Moreover, by the appropriate choice of the geometrical parameters
of the cavity one can excite different types of QBS (V-, W-,
$\Pi$-type, for example), allowing in this way diverse device
designing.

In summary, we have proposed the construction of frequency
selective switches and beam splitters using 2D multi-port waveguides.
Using ray as well as wave dynamics we shown that the switching and
splitting mechanism is based on tunneling into classically forbidden
phase space regions.
Finally, we want to stress that the choice of the cosine billiard
as waveguide cavity does not restrict the applicability of our results,
since the construction of frequency selective switches and beam splitters,
as described above, only requires a cavity characterized by an incomplete
horseshoe.

\bigskip
JAMB (antonio@chaos.gwdg.de) thanks support from the GIF,
the German-Israeli Foundation for Scientific Research and Development.
We thank J. P. Bird, R. Fleischmann and G. A. Luna-Acosta for useful discussions.

\pagebreak

\end{document}